%% file: main.tex
\documentclass{article}
\usepackage{spconf,amsmath,amsfonts,graphicx}
\usepackage{amsmath,amsfonts,tabularx,multirow}
\usepackage{amssymb,textcomp,mathtools}
\usepackage{color}
\usepackage{algorithm} 
\usepackage{algorithmic}  
\usepackage[algo2e]{algorithm2e} 
\usepackage{bm,mathtools,upgreek}
\usepackage{verbatim, cite}
\usepackage{bm,upgreek,algorithm,hyperref}
\usepackage{multirow, booktabs, hhline, array}
\usepackage{cite, url, makecell, setspace}
\usepackage{acronym}
\usepackage{xcolor}

\def\thline{\noalign{\hrule height 1.0pt}}

\renewcommand{\vec}[1]{\bm{\mathrm{#1}}}
\title{Real-time binaural speech separation with preserved spatial cues}
\name{Cong~Han, Yi~Luo, Nima~Mesgarani}
\address{Department of Electrical Engineering, Columbia University, New York, NY}

\begin{document}
\ninept
\maketitle
\setlength{\abovedisplayskip}{5pt}
\setlength{\belowdisplayskip}{5pt}
\setlength{\abovedisplayshortskip}{5pt}
\setlength{\belowdisplayshortskip}{5pt}
\begin{abstract}
\input{abstract}
\end{abstract}
\section{Introduction}
\label{sec:intro}
\input{introduction}

\section{MIMO TasNet for binaural speech separation}
\label{sec:bi}
\input{binaural.tex}

\section{Experimental settings}
\label{sec:exp}
\input{experiments}

\section{Results and discussions}
\label{sec:results}
\input{results}

\section{Conclusion}
\label{sec:con}
\input{conclusion}

\section{Acknowledgement}
\label{sec:ack}
This work was funded by a grant from the National Institute of Health, NIDCD, DC014279; a National Science Foundation CAREER Award.

\clearpage
\newpage
\bibliographystyle{IEEEbib}
\bibliography{Template}

\end{document}

%% file: abstract.tex
Deep learning speech separation algorithms have achieved great success in improving the quality and intelligibility of separated speech from mixed audio. Most previous methods focused on generating a single-channel output for each of the target speakers, hence discarding the spatial cues needed for the localization of sound sources in space. However, preserving the spatial information is important in many applications that aim to accurately render the acoustic scene such as in hearing aids and augmented reality (AR). Here, we propose a speech separation algorithm that preserves the interaural cues of separated sound sources and can be implemented with low latency and high fidelity, therefore enabling a real-time modification of the acoustic scene. Based on the time-domain audio separation network (TasNet), a single-channel time-domain speech separation system that can be implemented in real-time, we propose a multi-input-multi-output (MIMO) end-to-end extension of TasNet that takes binaural mixed audio as input and simultaneously separates target speakers in both channels. Experimental results show that the proposed end-to-end MIMO system is able to significantly improve the separation performance and keep the perceived location of the modified sources intact in various acoustic scenes.


%% file: introduction.tex
In real-world multi-talker acoustic environments, humans can easily separate speech and accurately perceive the location of each speaker due to the binaural acoustic features such as interaural time differences (ITDs) and interaural level differences (ILDs). Speech processing methods aimed to modify the acoustic scene are therefore required to not only separate sound sources, but do so in a way to preserve the spatial cues needed for accurate localization of sounds.

However, most of the binaural speech separation systems \cite{zhang2017deep, liu2018iterative, dadvar2019robust} are multi-input-single-output (MISO), and hence lose the interaural cues at the output level which are important for humans to perform sound lateralization and localization \cite{sams1993human, yin2002neural}. To achieve binaural speech separation as well as interaural cues preservation, the multi-input-multi-output (MIMO) setting is necessary, and currently, such setting can be divided into three main categories. 
 
The first category of methods add another stage for binaural sound rendering, such as head related transfer function (HRTF) hypotheses, after a MISO system \cite{zohourian2018gsc}. This method decouples speech separation and spatial cues preservation, however, it requires robust speaker localization algorithms and a priori knowledge about the HRTF of the listener \cite{algazi2001cipic}. Thus, it not only requires additional efforts but limits the system to be listener-dependent. 

The second category calculates a real-valued spectro-temporal mask and then applies the same mask to both left and right microphone channels \cite{lotter2006dual, rohdenburg2007robustness, reindl2010analysis, marin2011perceptually, azarpour2017binaural, zohourian2018gsc}. Because both sides obtain the same zero-phase gain, the original interaural cues are preserved. However, the separation performance may be limited because of the single-channel mask estimation and the constraint due to the same gain assumption.
 
In the third category, complex-valued filters are applied to all available microphone signals simultaneously to generate binaural outputs with an additional constraint on interaural cues preservation. One approach is to use two beamformers at the same time to generate left and right outputs respectively, such as generalized sidelobe canceller (GSC) \cite{gannot2001signal} and binaural minimum variance distortionless response (MVDR) beamformer \cite{hadad2015theoretical}. Another approach is multi-channel wiener filter (MWF) \cite{doclo2006theoretical} that is equivalent to the combination of spatial filtering and spectral post-filtering. There has been a method that exploits the deep neural network to estimate complex ideal ratio masks (cIRM) for both left and right channels \cite{sun2019deep}. 
Since these multi-channel methods aim at estimating the desired separated sources in each channel, the spatial information could be naturally preserved.
 
One common issue for the systems mentioned above is that the system latency can be perceivable by humans, and the delayed playback of the separated speakers might affect the localization of the signals due to the precedence effect \cite{haas1972influence}. To decrease the system latency while maintaining the separation quality, a natural way is to use time-domain separation methods with smaller windows. Recent deep learning-based time-domain separation systems have proven their effectiveness in achieving high separation quality and decreasing the system latency \cite{stoller2018wave, venkataramani2018end, luo2019conv, zhang2020furcanext}, however, all those systems are still MISO and their ability to perform binaural speech separation and interaural cues preservation is not fully addressed. 

In this paper, we look into multiple methods for formulating such systems into MIMO systems and investigate their capability of high-quality separation and interaural cue preservation. Based on the time-domain audio separation network (TasNet) \cite{luo2019conv}, we propose a MIMO TasNet that takes binaural mixture signals as input and simultaneously separates speech in both channels, then the separated signals can be directly rendered to the listener without post-processing. The MIMO TasNet exploits a parallel encoder to extract cross-channel information for mask estimation and uses mask-and-sum method to perform spatial and spectral filtering for better separation performance. We compare it with other variants of TasNet in the tasks. Experiment results show that MIMO TasNet can perform listener-independent speech separation across a wide range of speaker angles and preserve both ITD and ILD features with significantly higher quality than the single-channel baseline. Moreover, the minimum system latency of the systems can be less than 5 ms, showing the potentials for the actual deployment of such systems into real-world hearable devices.

The rest of the paper is organized as follows. We introduce the problem definition of binaural separation with preserved spatial cues and the MIMO variants of TasNet in Section~\ref{sec:bi}, describe the experiment settings in Section~\ref{sec:exp}, discuss the results in Section~\ref{sec:results}, and concludes the paper in Section~\ref{sec:con}.

\begin{figure*}[ht]
    \small
    \centering
    \includegraphics[width=2\columnwidth]{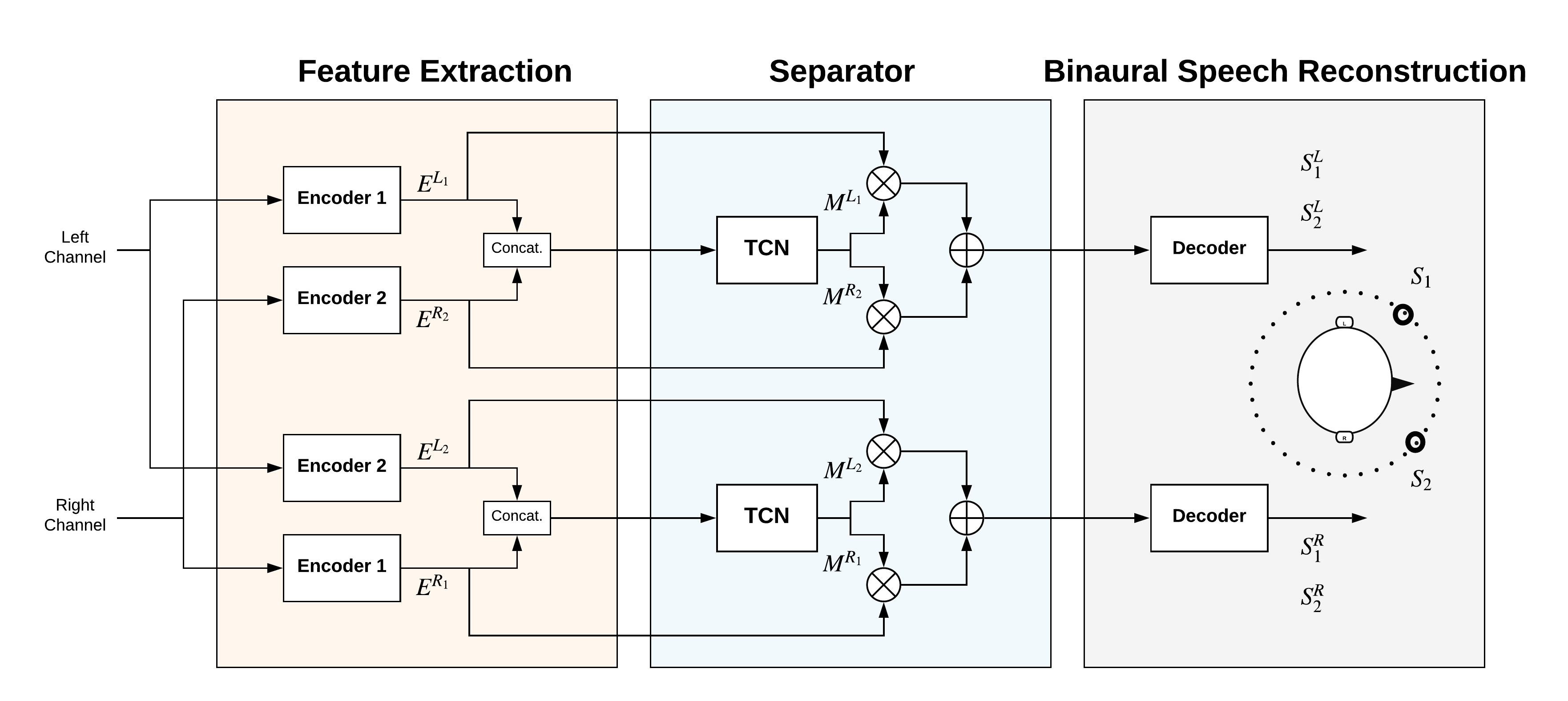}
    \caption{The architecture of the proposed binaural speech separation network. Two encoders are shared by the mixture signals from both channels, and the encoder outputs for each channel are concatenated together and passed to a mask estimation network. Then, spectral-temporal and spatial filtering are performed by applying the masks to the corresponding encoder outputs and sum them up on both left and right paths. Finally, binaural separated speech are reconstructed by a linear decoder.}
    \label{fig:arch}
\end{figure*}

%% file: binaural.tex
\subsection{Problem definition}
\label{sec:formula}
The problem of binaural speech separation is formulated as the separation of $C$ sources $\vec{s}_i^{l,r}(t) \in \mathbb{R}^{1\times T}, \, i=1,\ldots,C$ from the binaural mixtures $\vec{x}^l(t), \vec{x}^r(t) \in \mathbb{R}^{1\times T}$, where the superscripts $l$ and $r$ denote the left and right channels, respectively. For preserving the interaural cues in the outputs, we consider the case where every single source signal is transformed by a set of head-related impulse response (HRIR) filters for a specific listener:
\begin{align}
    \begin{cases}
    \vec{s}_i^l = \hat{\vec{s}}_i \circledast \vec{h}_i^l \\
    \vec{s}_i^r = \hat{\vec{s}}_i \circledast \vec{h}_i^r \\
    \end{cases} \, i = 1, \ldots, C
\end{align}
where $\hat{\vec{s}}_i \in \mathbb{R}^{1\times T'}$ is the monaural signal of source $i$, $\vec{h}_i^l, \vec{h}_i^r \in \mathbb{R}^{1\times (T-T'+1)}$ are the pair of HRIR filters corresponding to the source $i$, and $\circledast$ represents the convolution operation. Using the HRIR-transformed signals as the separation targets forces the model to preserve interaural cues introduced by the HRIR filters, and the outputs can be directly rendered to the listener.

\subsection{MIMO TasNet}
\label{sec:single}

\subsubsection{TasNet overview}
\label{sec:tasnet}

TasNet has been shown to achieve superior separation performance in single-channel mixtures \cite{luo2019conv}. TasNet contains three modules: a linear encoder first transforms the mixture waveform into a two-dimensional representation similar to spectrograms; a separator estimates $C$ multiplicative functions similar to time-frequency masks based on the 2-D representation; and a linear decoder transforms the $C$ target source representations back to waveforms.

Various approaches have been proposed to extend TasNet into the multi-channel framework \cite{gu2019end, luo2019fasnet}. A standard pipeline is to incorporate cross-channel features into the single-channel model, where spatial features such as interaural phase difference (IPD) is concatenated with the mixture encoder output on a selected reference microphone for mask estimation \cite{gu2019end}. In various scenarios, such configuration have led to a significantly better separation performance than the signal-channel TasNet.

\subsubsection{Design of MIMO TasNet}
\label{sec:parallel}

The proposed MIMO TasNet uses a parallel encoder for spectro-temporal and spatial features extraction and a mask-and-sum mechanism for source separation. A \textit{primary} encoder is always applied to the channel to be separated, and a \textit{secondary} encoder is applied to the other channel to jointly extract cross-channel features. In other words, the sequential order of the encoders determines which channel (left of right) the separated outputs belong to. The outputs of the two encoders are concatenated and passed to the separator, and $2C$ multiplicative functions are estimated for the $C$ target speakers. $C$ multiplicative functions are applied to the \textit{primary} encoder output while the other $C$ multiplicative functions are applied to the \textit{secondary} encoder output,  and the two multiplied results are then summed to create representations for $C$ separated sources. We denote it as the \textit{mask-and-sum} mechanism to distinguish it from the other methods where only $C$ multiplicative functions were estimated from the separation module and applied to only the reference channel. Similar to TasNet, a linear decoder transforms the $C$ target source representations back to waveforms.  Figure~\ref{fig:arch} shows the flowchart of the system design.

Note that a parallel encoder design for multi-channel TasNet has been discussed in a previous literature \cite{gu2019end}. For a $N$-channel input, $N$ encoders were applied to each of them and the encoder outputs were summed to create a single representation. The multiplicative function was also estimated on the single representation, which resulted in a MISO system design. We can easily find that it is a special case of MIMO TasNet where the two multiplicative functions for the two encoders are equal. Although in \cite{gu2019end} an on par performance with respect to the feature concatenation method was reported for the parallel encoder design, in Section~\ref{sec:results} we will show that MIMO TasNet is able to significantly surpass feature concatenation TasNet in various configurations in both separation performance and spatial cue preservation accuracy.

\subsubsection{Training objective}
\label{sec:obj}

Scale-invariant signal-to-distortion ratio (SI-SDR) \cite{le2019sdr} is used as both the evaluation metric and training objective for many recent end-to-end separation systems. SI-SDR between a signal $\vec{x} \in \mathbb{R}^{1\times T}$ and its estimate $\hat{\vec{x}} \in \mathbb{R}^{1\times T}$ is defined as:
\begin{align}
\label{eqn:sisdr}
    \text{SI-SDR}(\vec{x}, \hat{\vec{x}}) = 10\,\text{log}_{10}\left(\frac{||\alpha\vec{x}||_2^2}{||\hat{\vec{x}} - \alpha\vec{x}||_2^2}\right)
\end{align}
where $\alpha = \hat{\vec{x}}\vec{x}^\top / \vec{x}\vec{x}^\top$ corresponds to the rescaling factor. Although SI-SDR is able to implicitly incorporate the ITD information, the scale-invariance property of SI-SDR makes it insensitive to power rescaling of the estimated signal, which may fail in preserving the ILD between the outputs. Hence instead of using SI-SDR as the training objective, we use the plain signal-to-noise ratio (SNR) defined as:
\begin{align}
\label{eqn:snr}
    \text{SNR}(\vec{x}, \hat{\vec{x}}) = 10\,\text{log}_{10}\left(\frac{||\vec{x}||_2^2}{||\hat{\vec{x}} - \vec{x}||_2^2}\right)
\end{align}

%% file: experiments.tex
\subsection{Dataset}

We generated an anechoic speech dataset from the WSJ0-2mix dataset \cite{hershey2016deep}. 30 hours of training data, 10 hours of validation data and 5 hours of test data were generated with the same configuration as the single-channel WSJ0-2mix data, while the clean speech are convolved with randomly sampled HRIR filters from the CIPIC HRTF Database \cite{algazi2001cipic}. The CIPIC HRTF Database contains real-recorded HRIR filters for 20 subjects across 25 different interaural-polar azimuths from $-80^{\circ}$ to $80^{\circ}$ and 50 different interaural-polar elevations from $-90^{\circ}$ to $270^{\circ}$. We randomly sampled two speaker locations in the database for spatial rendering. We used 27 subjects for training and validation sets and 9 unseen subjects for test set, ensuring that the model is evaluated in a listener-independent way. All mixtures were downsampled to 8k Hz.

Anechoic WSJ0-3mix dataset with spatial cues was generated by using the the same method as above. To generate noisy WSJ0-2mix dataset, we added to the training set the noise from one out of eight environmental noises (washing room, kitchen, sport field, city park, office, meeting room) chosen from DEMAND dataset\cite{thiemann2013demand} with SNR between -15 and 2.5 dB. The noise in test set is from another eight scenarios (subway station, restaurant, public square, traffic intersection, subway, private car). To generate echoic WSJ0-2mix dataset, HRIR filters are obtained from the BRIR Sim Set\footnote{\url{http://iosr.uk/software/index.php}}, which is simulated with different reverberation time (T60). We use rooms with T60 0.1s, 0.2s, 0.4s, 0.5s, 0.7s, 0.8s, 1.0s for training and 0.3s, 0.6s, 0.9s for testing.

\subsection{Evaluation metrics}

We evaluate the model with both the separation quality and the ability to preserve interaural cues. SNR improvement (SNRi) is used as the signal quality metric instead of SI-SDR improvement according to our discussion in Section~\ref{sec:obj}. ITD and ILD errors between the separated and target clean signals are used as the metric for the accuracy of preserving interaural cues, which are defined as:
\begin{align}
\Delta_{ITD} &= \left|\rm{ITD}(\vec{s}^l, \vec{s}^r) - \rm{ITD}(\bar{\vec{s}}^l, \bar{\vec{s}}^r)\right| \\
\Delta_{ILD} &= \left|10\log_{10} \frac{||\vec{s}^l||_2^2}{||\vec{s}^r||_2^2} - 10\log_{10} \frac{||\bar{\vec{s}}^l||_2^2}{||\bar{\vec{s}}^r||_2^2}\right|
\end{align}
where $\bar{\vec{s}}^l, \bar{\vec{s}}^r \in \mathbb{R}^{1\times T}$ are the separated signals in left and right channels, respectively, $\vec{s}^l, \vec{s}^r \in \mathbb{R}^{1\times T}$ are the corresponding target signals, and $||\cdot||$ denotes the $L_2$-norm of the signal. We use generalized cross-correlation phase transform (GCC-PHAT) algorithm \cite{knapp1976generalized} to compute time difference of arrival (TDOA) of $\vec{s}^l$ and $\vec{s}^r$ as $\rm{ITD}(\vec{s}^l, \vec{s}^r)$. The tool is available online \footnote{\url{https://www.mathworks.com/help/phased/ref/gccphat.html}}.

\subsection{Network architectures}
The configurations of the MIMO TasNet variants are based on the causal setting of the single-channel TasNet \cite{luo2019conv}. We use 64 filters in the linear encoder and decoder with 2 ms filter length (i.e. 16 samples at 8k Hz). In the causal temporal convolutional network (TCN), there are 4 repeated stacks and each one includes 8 1-D convolutional blocks. The number of parameters in all models are aligned to 1.67M for a fair comparison. 

For baseline models, we adopt the following configurations:
\begin{enumerate}
    \item \textit{Single-channel TasNet}: the single-channel model is applied to each channel independently.
    \item \textit{Feature concatenation TasNet}: cross-channel features are concatenated to the encoder output in the same way as \cite{gu2019end}. We use sin(IPD), cos(IPD) and ILD as spatial features, where the IPD and ILD are defined as
    \begin{align}
        &\rm{IPD}(\vec{X}, \vec{Y}) = \angle\vec{X} -  \angle\vec{Y} \\
        &\rm{ILD}(\vec{X}, \vec{Y}) = 10\log_{10}\left( |\vec{X}| \oslash |\vec{Y}| \right)
    \end{align}
    where $\vec{X}, \vec{Y}$ are the spectrograms of the two channel mixtures, $\oslash$ means element-wise division. The window length of STFT for calculating spectrograms is 256 samples.
    \item \textit{Parallel encoder TasNet}: the same configuration as in \cite{gu2019end} which is also discussed in Section~\ref{sec:parallel}.
\end{enumerate}

%% file: results.tex
\begin{table*}[!ht]
	\small
	\centering
	\caption{SNR improvement (dB), ITD error ($\mu$s), and ILD error (dB) for different variants of TasNet on anechoic spatialized WSJ0-2mix. The averaged performance on different ranges of speaker angles is reported.}
	\label{tab:wsj0-2mix}
	\begin{tabular}{c|ccccc}
		\thline
        \multirow{3}{*}{\thead{Method}} &  \multicolumn{5}{c}{\thead{SNRi / $\Delta_{ITD}$ / $\Delta_{ILD}$}} \\
        \cline{2-6}
        & \multicolumn{5}{c}{\thead{Angle}} \\
        & $<$15\textdegree & 15-45\textdegree & 45-90\textdegree & $>$90\textdegree & Average \\
        \thline
        \multicolumn{1}{l|}{TasNet}
        & 10.0 / 6.0 / {\bf 0.29} 
        & 10.0 / 5.8 / 0.39	
        & 10.3 / 4.8 / 0.56	
        & 10.7 / 6.4 / 0.59	
        & 10.2 / 5.8 / 0.46 \\
        \hline
        \multicolumn{1}{l|}{+ILD}
        & 11.1 / 3.1 / 0.31	
        & 13.4 / 1.9 / 0.12	
        & 14.4 / 1.3 / 0.16	
        & 14.8 / 1.9 / 0.17	
        & 13.4 / 2.0 / {\bf 0.19} \\
        \hline
        \multicolumn{1}{l|}{+sin(IPD), cos(IPD)}
        & 11.7 / {\bf 2.4} / 0.34 
        & 14.1 / 1.7 / 0.14 
        & 14.7 / 1.4 / 0.20 
        & 15.3 / 2.0 / 0.20 
        & 13.9 / 1.9 / 0.22 \\
        \hline
        \multicolumn{1}{l|}{+sin(IPD), cos(IPD), ILD}	
        & {\bf 11.8} / {\bf 2.4} / 0.33 
        & 14.5 / 1.6 / {\bf 0.11} 
        & 15.3 / 1.2 / 0.16 
        & 15.8 / {\bf 1.8} / 0.18 
        & 14.4 / {\bf 1.8} / 0.20 \\
        \hline
        \multicolumn{1}{l|}{+parallel encoder}
        & 10.6 / 3.0 / 0.47	
        & 15.1 / 1.5 / {\bf 0.11}
        & 16.8 / 1.2 / 0.11
        & 17.7 / 2.0 / 0.12
        & 15.0 / 2.0 / 0.20 \\
		\hline
		\multicolumn{1}{l|}{+parallel encoder, mask\&sum}
        & 10.7 / 2.8 / 0.47 
        & {\bf 15.6} / {\bf 1.3} / 0.13  
        & {\bf 17.7} / {\bf 1.1} / {\bf 0.09 } 
        & {\bf 18.3} / {\bf 1.8} / {\bf 0.09 }
        & {\bf 15.6} / {\bf 1.8} / {\bf 0.19 } \\
		\thline
    \end{tabular}
    \label{tab:t1}
\end{table*}

\begin{table*}[!ht]
	\small
	\centering
	\caption{Evaluation of TasNet with parallel encoder on several adverse conditions: three-speaker separation, two-speaker separation with environmental noise, and with room reverberance.}
	\label{tab:wsj0-2mix}
	\resizebox{\textwidth}{!}
	{
	\begin{tabular}{c|c|ccc|ccc}
		\thline
        \multirow{3}{*}{\thead{Method}} &  \multicolumn{7}{c}{\thead{SNRi / $\Delta_{ITD}$ / $\Delta_{ILD}$}} \\
        \cline{2-8}
        & \multicolumn{1}{c|}{\thead{3 speaker}} & \multicolumn{3}{c|}{\thead{2 speaker with noise (SNR)}} & \multicolumn{3}{c}{\thead{2 speaker with reverberance (RT60)}} \\
        & & 12.5 dB & 5 dB & -2.5 dB & 0.3s & 0.6s & 0.9s \\
        \thline
        \multicolumn{1}{l|}{TasNet}
        & 9.1 / 6.3 / 0.74 
        & 9.8 / 3.4 / 0.31	
        & 10.9 / 3.7 / 0.31 
        & 13.8 / 5.2 / 0.57 	
        & 7.2 / 10.8 / 0.46 
        & 6.2 / 45.1 / 0.47 
        & 5.7 / 44.5 / 0.50 \\
        \hline
        \multicolumn{1}{l|}{+parallel encoder}
        & 11.3 / 12.3 / 0.84 
        & 13.7 / 2.3 / 0.16 
        & 15.0 / 2.5 / 0.18 
        & 17.8 / 3.0 / 0.23 
        & 9.2 / 6.5 / {\bf 0.20}  
        & 7.7 / 33.2 / 0.25  
        & 6.9 / 17.7 / 0.30 \\
		\hline
		\multicolumn{1}{l|}{+parallel encoder, mask\&sum}
        & {\bf 12.1} / {\bf 5.7} / {\bf 0.45}
        & {\bf 14.3} / {\bf 2.2} / {\bf 0.14} 
        & {\bf 15.3} / {\bf 2.3} / {\bf 0.15}
        & {\bf 18.2} / {\bf 2.8} / {\bf 0.21} 
        & {\bf 9.4} / {\bf 5.9} / 0.23 
        & {\bf 7.8} / {\bf 30.0} / {\bf 0.21}
        & {\bf 7.1} / {\bf 15.6} / {\bf 0.25} \\
		\thline
    \end{tabular}
    }
    \label{tab:t2}
\end{table*}

Table~\ref{tab:wsj0-2mix} compares different MIMO TasNet variants at various speaker locations on anechoice spatialized WSJ0-2mix. The single-channel baseline is able to achieve the smallest ILD error across all models when the speaker angle is very small, which indicates that the interaural features in this scenario are not helpful in preserving the absolute energy of the separated speech. For all other speaker locations, both the ILD error and separation quality for the single-channel model are significantly worse than all the MIMO variants. For TasNet concatenated with sin(IPD), cos(IPD) and ILD features, we can observe significant signal quality improvement and ITD/ILD error reduction across all angle ranges, and better performance is achieved with larger speaker angle. This confirms the previous observations regarding the effectiveness of cross-channel features in end-to-end frameworks \cite{gu2019end}. The parallel encoder method has on par performance in preserving ITD/ILD with feature concatenation method, but achieves better separation performance except when the speaker angle is small (less than 15\textdegree). The significant improvement for signal quality (SNRi) indicates that the parallel encoders are able to implicitly extract more effective cross-channel features than cross-domain features IPD/ILD for multi-channel speech separation. The further improvement from mask-and-sum mechanism indicates the effectiveness of combining spatial filtering and spectral filtering to separate sources. The correlation (Pearson’s r) between SNRi and $\Delta_{ITD}$ and between SNRi and $\Delta_{ILD}$ are -0.77 and -0.85, respectively (p \textless 0.0001 for both), which means higher separated signal quality helps in preserving ITD/ILD better.

To further examine our proposed MIMO TasNet in more adverse environments, we tested the separation accuracy in three speaker mixtures, noisy speech separation and speech separation with room reverberation. Note that in the evaluation of these three cases, top 5\% $\Delta_{ITD}$ and $\Delta_{ILD}$ were dropped before averaging to prevent the errors incurred by outliers. 

When testing the model on noisy WSJ0-2mix dataset, we set the noise power range at three levels. As shown in Table~\ref{tab:t2}, additive noise contaminates both speech quality and ITD/ILD preservation, but the overall performance compared to the clean condition is still superior and MIMO TasNet with parallel encoder and mask-and-sum achieves the best performance in all metrics across all noise levels, which proves the MIMO TasNet is more robust to the noise. 

We observe that three speaker separation is more challenging than noisy speech separation. Both ITD and ILD preservation downgrade significantly than the two-speaker case. That's because the model had failed to separate some of speech with very small power compared to the other two speakers or speech with very similar spatial features to others, and the failure of separation leads to the failure of ITD/ILD preservation.

Finally, we evaluated the model on the echoic spatialized WSJ0-2mix dataset. The target is the reverberant clean signal. Not surprisingly, convolutive room reverberation is a more challenging condition than additive environmental noises both in terms of signal quality improvement and preserving spatial cues as the sparseness properties of the speech is affected by room reverberation. The smearing casued by reverberation means that the mixture at each instance includes components of the same and different speakers, which makes the mask prediction and TDOA estimation more difficult. As a result, SNRi and $\Delta_{ITD}$ are more easily affected by the reverberation. Also, using only two channels doesn't fully take advantage of multi-channel algorithms to reduce the influence of reverberation. Nonetheless, averaged 9.4 dB SNR improvement, 5.9 $\mu s$ ITD error and 0.23 dB ILD error shows that the performance of MIMO TasNet is still helpful in the moderate reverberant environment.

%% file: conclusion.tex
In this paper, we investigated the problem of real-time binaural speech separation with interaural cues preservation. We proposed a multi-input-multi-output (MIMO) TasNet that uses a parallel encoder and mask-and-sum mechanism to improve performance. Experimental results show that the MIMO TasNet is able to achieve significantly better separation performance and has the ability to preserve interaural time difference (ITD) and interaural level difference (ILD) features in the separated outputs compared to the other existing variants of TasNet.  Future works include adapting to environmental noise and room reverberation and incorporate extra microphones for obtaining more cross-channel information, which can pave the way to real-world speech separation solutions for acoustic scene modification.